\documentclass[english,a4paper,manuscript]{revtex4}
\usepackage[T1]{fontenc}
\usepackage[latin1]{inputenc}
\usepackage{amsmath}

\makeatletter

\makeatletter

\usepackage{geometry}

\makeatother

\usepackage{babel}
\makeatother
\begin{document}

\title{Relativistic transport theory for simple fluids to first order in
the gradients}

\author{A. Sandoval-Villalbazo$^{1}$, A. L. Garcia-Perciante$^{2}$, L.
S. Garcia-Colin$^{3}$ }

\address{$^{1}$Depto. de Fisica y Matematicas, Universidad Iberoamericana,
Prolongacion Paseo de la Reforma 880, Mexico D. F. 01219, Mexico.
$^{2}$Depto. de Matematicas Aplicadas y Sistemas, Universidad Autonoma
Metropolitana-Cuajimalpa, Artificios 40 Mexico D.F 01120, Mexico.
$^{3}$Depto. de Fisica, Universidad Autonoma Metropolitana-Iztapalapa,
Av. Purisima y Michoacan S/N, Mexico D. F. 09340, Mexico. Also at
El Colegio Nacional, Luis Gonzalez Obregon 23, Centro Historico, Mexico
D. F. 06020, Mexico}

\begin{abstract}
In this paper we show how using a relativistic kinetic equation the
ensuing expression for the heat flux can be casted in the form required
by Classical Irreversible Thermodynamics. Indeed, it is linearly related
to the temperature and number density gradients and not to the acceleration
as the so called ¨first order in the gradients¨ theories propose.
Since the specific expressions for the transport coefficients are
irrelevant for our purposes, the BGK form of the kinetic equation
is used. Moreover, from the resulting hydrodynamic equations it is
readily seen that the equilibrium state is stable in the presence
of the spontaneous fluctuations in the transverse hydrodynamic velocity
mode of the simple relativistic fluid. The implications of this result
are thoroughly discussed.
\end{abstract}
\maketitle

\section{Introduction}

In his seminal paper on classical irreversible relativistic thermodynamics,
C. Eckart in 1949 \cite{Eckart} introduced a bold assumption which
has had a profound influence on the subject. Indeed, in order to account
for dissipative effects in a non-ideal (viscous) fluid he introduced
the heat flow present in the fluid as part of Einstein's matter-stress
tensor $\mathcal{T}_{\ \mu\nu}$. Being heat a non-mechanical form
of energy, this has raised a rather strong debate about the physical
meaning of such an assumption \cite{JNT06}-\cite{JNT07}. Moreover,
his formalism often referred to as a \textit{first order theory} \cite{israel,israel-stewart}
has given rise to additional controversies. In a classical work Hiscock
and Lindblom \cite{HL} showed that in the resulting linearized version
of the hydrodynamic equations, the fluctuations in the transverse
mode of the hydrodynamic velocity grow exponentially with time. This
is a consequence of the fact that Eckart, in view of his assumption,
was rather forced to introduce a constitutive equation for the heat
flow $J_{[Q]}^{\alpha}$ given by the equation \begin{equation}
J_{[Q]}^{\alpha}=-\kappa\left(h^{\alpha\beta}T_{,\beta}-\frac{T}{c^{2}}\dot{u}^{\alpha}\right)\label{uno}\end{equation}
where $\kappa$ is the heat conductivity, $T$ is the local temperature,
$\ h_{\beta}^{\alpha}=\delta_{\beta}^{\alpha}+\frac{1}{c^{2}}u^{\alpha}u_{\beta}$
is the spatial projector, $u^{\alpha}$ is the hydrodynamic velocity
and $\dot{u}^{\alpha}$ its proper time derivative. Also $c$ is the
speed of light and $u^{\alpha}u_{\alpha}$ $=-c^{2}$. Notice further
that Eq. (\ref{uno}) is at odds with the basic tenets of linear irreversible
thermodynamics (LIT). In fact, this formalism requires that constitutive
equations linearly relate the forces with the fluxes \cite{meix,dgm}.
This requirement is met by the first term in Eq. (\ref{uno}) but
certainly not by the second where neither $T/c^{2}$ nor $\dot{u}^{\alpha}$
represent a thermodynamic force or a flux. Further, physical insight
of this feature of Eckart's formalism has recently been discussed
elsewhere \cite{Nos1}.

The purpose of this work is to show that Eq. (\ref{uno}) does not
necessarily follow from a relativistic kinetic theory treatment. Moreover,
we establish an expression relating the heat flux and its thermodynamic
forces, \emph{to first order in the gradients}, that leads to an exponential
decay in the spontaneous fluctuations for the transverse hydrodynamic
velocity mode. To accomplish this task in Sect. II we derive the set
of linearized hydrodynamic equations which result from a relativistic
Boltzmann equation of the BGK type, the so-called Marle equation,
and in Sect. III we discuss the physical implications that may be
drawn from such set.

\section{Linearized Relativistic Hydrodynamics}

To obtain the above mentioned set of linear relativistic equations
for a simple viscous fluid we start by considering the special relativistic
Boltzmann equation in the absence of external forces: \begin{equation}
v^{\alpha}f_{,\alpha}=J\left(f,\, f^{\prime}\right)\label{eq:1}\end{equation}
The term on the right hand side of Eq. (\ref{eq:1}) is the collision
operator which in general is a bilinear form containing the products
of the distribution function $f$ evaluated before and after a collision
takes place. When one introduces the probability for the occurrence
of each one of these binary collisions, namely the cross section and
integrates over all possible velocities for the particles, one obtains
the general Boltzmann equation. Here, for the sake of simplicity,
the kernel is modeled by means of the BGK approximation \cite{Liboff}\cite{CK},
namely\begin{equation}
J\left(f,\, f^{\prime}\right)=-\frac{f-f^{\left(0\right)}}{\tau}\label{eq:q}\end{equation}
This approximation consists in replacing all the details of such binary
collisions by the relaxation time $\tau$, which can be viewed then
as an adjustable parameter in the theory. This assumption simplifies
the ensuing arguments for which the full collisional term adds no
significant changes. The molecular four velocity, $v^{\alpha}$ is

\noindent \begin{equation}
v^{\alpha}=\left[\begin{array}{c}
\gamma w^{\ell}\\
\gamma c\end{array}\right]\label{eq:1.1}\end{equation}
where $w^{\ell}$ is the molecular three-velocity and $\gamma=\left(1-w^{\ell}w_{\ell}/c^{2}\right)^{-1/2}$
is the usual relativistic factor. All greek indices run from 1 to
4 and the latin ones run up to 3. The local variables are the weighted
averages of the microscopic quantities. In this context, the number
density reads

\noindent \begin{equation}
n=\int f\gamma dv^{*}\,,\label{eq:01}\end{equation}
 and the thermodynamic average of a dynamical quantity $\psi$ is
defined through: \begin{equation}
\left\langle \psi\right\rangle =\frac{1}{n}\int\gamma\psi fdv^{*}\label{prom}\end{equation}
 with $dv^{*}=\gamma^{5}\frac{cd^{3}w}{v^{4}}$ \cite{Liboff}. Now,
standard kinetic theory leads to the energy balance equation \cite{CK,degrootrel}:
\begin{equation}
\frac{\partial}{\partial t}\left(n\left\langle mc^{2}\gamma\right\rangle \right)+\left(n\left\langle mc^{2}\gamma w^{\ell}\right\rangle \right)_{;\ell}=0\label{Baley}\end{equation}
 In this expression one can easily identify the heat flux as: \begin{equation}
J_{[Q]}^{\ell}=n\left\langle mc^{2}v^{\ell}\right\rangle =mc^{2}\int v^{\ell}f\gamma dv^{*}\label{Baley2}\end{equation}
which vanishes if the average is calculated using the equilibrium
distribution function. 

\noindent At the Navier-Stokes level, following a Chapman-Enskog expansion,
the distribution function can be written as \begin{equation}
f=f^{\left(0\right)}\left(1+\phi\right)\label{eq:3}\end{equation}
where $\phi$ is the first order correction in the Knudsen parameter,
a weighted measure of the gradients in the system. For particles of
rest mass $m$, relativistic parameter $z=\frac{kT}{mc^{2}}$ and
in the non-degenerate case, the equilibrium function reads \cite{CK,degrootrel}:

\begin{equation}
f^{\left(0\right)}=\frac{n}{4\pi c^{3}z\mathcal{K}_{2}\left(\frac{1}{z}\right)}\text{Exp}\left(\frac{u^{\beta}v_{\beta}}{zc^{2}}\right)\,.\label{eq:4}\end{equation}
in which $k$ is Boltzmann's constant, and $\mathcal{K}_{n}$ is the
modified Bessel function of the $n$-th kind. Notice also that, if
the fluid is at rest, $u^{\beta}v_{\beta}=-\gamma c^{2}$.

The function $\phi$ is obtained by a simple procedure. Direct substitution
of Eq. (\ref{eq:3}) in Eq. (\ref{eq:1}) leads to \begin{equation}
\phi=-\tau v^{\alpha}f_{,\alpha}^{\left(0\right)}=-\tau v^{\alpha}\left(\frac{\partial f^{\left(0\right)}}{\partial n}n_{,\alpha}+\frac{\partial f^{\left(0\right)}}{\partial T}T_{,\alpha}+\frac{\partial f^{\left(0\right)}}{\partial u^{\beta}}u_{;\alpha}^{\beta}\right)\label{eq:5}\end{equation}
where the second equality is due to the fact that the distribution
function is a time dependent functional of the local variables $n$,
$T$ and $u^{\nu}$. The derivatives in Eq. (\ref{eq:5}) are given
by \begin{equation}
\frac{\partial f^{\left(0\right)}}{\partial n}=\frac{f^{\left(0\right)}}{n}\label{eqq}\end{equation}
 \begin{equation}
\frac{\partial f^{\left(0\right)}}{\partial T}=\left(-1+\frac{\gamma}{z}-\frac{\mathcal{K}_{1}\left(\frac{1}{z}\right)}{2z\mathcal{K}_{2}\left(\frac{1}{z}\right)}-\frac{\mathcal{K}_{3}\left(\frac{1}{z}\right)}{2z\mathcal{K}_{2}\left(\frac{1}{z}\right)}\right)\frac{f^{\left(0\right)}}{T}\label{eq8}\end{equation}
and \begin{equation}
\frac{\partial f^{\left(0\right)}}{\partial u^{\beta}}=\frac{v_{\beta}}{zc^{2}}\, f^{\left(0\right)}\label{eq9}\end{equation}
Notice that a hydrodynamic acceleration contribution appears in the
last term of Eq. (\ref{eq:5}) for $\alpha=4$. If this term is left
untouched and the heat flow is calculated with Eq. (\ref{Baley2})
one is led to the difficulty inherent in Eq. (\ref{uno}). On the
other hand, according to the prescription of Hilbert-Enskog's method
\cite{Liboff}-\cite{degrootrel} this acceleration term is eliminated
using Euler's equations, which precisely relate the acceleration to
the pressure gradient. Moreover, this method establishes that by choosing
the local variables, $n$, $T$ and $u^{\nu}$ defined only through
$f^{\left(0\right)}$, one guarantees the uniqueness of the solution
given in Eq. (\ref{eq:3}). Consistently with the fact that we have
chosen $n$, $T$ and $u^{\nu}$ as the state variables describing
the local thermodynamic states of the fluid, $\nabla p$ has to be
expressed in terms of them. This is accomplished using the local equilibrium
assumption which allows expressing $\nabla p$ in terms of $\nabla T$
and $\nabla n$ in complete agreement with the tenets of classical
irreversible thermodynamics namely,\begin{equation}
\nabla p=\left(\frac{\partial p}{\partial T}\right)_{n}\nabla T+\left(\frac{\partial p}{\partial n}\right)_{T}\nabla n\label{eq:sand}\end{equation}
 Finally, since for a simple inviscid fluid whose dynamics is described
by Eq. (\ref{eq:1}), the equation of state $p=nkT$ holds true locally
\cite{CK}, a straightforward calculation leads to the following form
for the heat flux \begin{equation}
J_{[Q]}^{\ell}=-L_{T}\frac{T^{,\ell}}{T}-L_{n}\frac{n^{,\ell}}{n}\label{eq:10}\end{equation}
Equation (\ref{eq:10}) is of the canonical form required by LIT where
the thermodynamic forces are the temperature and the density gradients,
respectively. The transport coefficients are given by\begin{eqnarray}
L_{T} & = & \tau\frac{4\pi}{3}mc^{7}\left\{ \left(\frac{1}{z}-\frac{p}{zc^{2}\left(\frac{n\varepsilon}{c^{2}}+\frac{p}{c^{2}}\right)}\right)\int f^{\left(0\right)}\gamma^{2}\left(\gamma^{2}-1\right)^{3/2}d\gamma\right.\nonumber \\
 &  & \left.-\left(1+\frac{\mathcal{K}_{1}\left(\frac{1}{z}\right)}{2z\mathcal{K}_{2}\left(\frac{1}{z}\right)}+\frac{\mathcal{K}_{3}\left(\frac{1}{z}\right)}{2z\mathcal{K}_{2}\left(\frac{1}{z}\right)}\right)\int f^{\left(0\right)}\gamma\left(\gamma^{2}-1\right)^{3/2}d\gamma\right\} \label{eq:11}\end{eqnarray}
 and\begin{equation}
L_{n}=\tau\frac{4\pi}{3}mc^{7}\left\{ \int f^{\left(0\right)}\gamma\left(\gamma^{2}-1\right)^{3/2}d\gamma-\frac{p}{zc^{2}\left(\frac{n\varepsilon}{c^{2}}+\frac{p}{c^{2}}\right)}\int f^{\left(0\right)}\gamma^{2}\left(\gamma^{2}-1\right)^{3/2}d\gamma\right\} \label{eq:12}\end{equation}
where $\varepsilon$ is the internal energy per particle. The integrals
in Eqs. (\ref{eq:11}) and (\ref{eq:12}) can be calculated in the
comoving frame. Thus, the transport coefficients can be written as\begin{equation}
L_{T}=nmc^{4}z^{2}\tau f_{T}\left(z\right)\qquad\qquad L_{n}=nmc^{4}z^{2}\tau f_{n}\left(z\right)\label{eq:L}\end{equation}
where the functions $f_{T}$ and $f_{n}$ are given by\begin{eqnarray}
f_{T}\left(z\right) & = & \left(\frac{1}{z}-\left(4z+\frac{\mathcal{K}_{1}\left(\frac{1}{z}\right)}{\mathcal{K}_{2}\left(\frac{1}{z}\right)}\right)^{-1}\right)\left(\frac{1}{z}+5\frac{\mathcal{K}_{3}\left(\frac{1}{z}\right)}{\mathcal{K}_{2}\left(\frac{1}{z}\right)}\right)\nonumber \\
 & - & \left(1+\frac{\mathcal{K}_{1}\left(\frac{1}{z}\right)}{2z\mathcal{K}_{2}\left(\frac{1}{z}\right)}+\frac{\mathcal{K}_{3}\left(\frac{1}{z}\right)}{2z\mathcal{K}_{2}\left(\frac{1}{z}\right)}\right)\frac{\mathcal{K}_{3}\left(\frac{1}{z}\right)}{z\mathcal{K}_{2}\left(\frac{1}{z}\right)}\label{eq:ep}\end{eqnarray}
 \begin{equation}
f_{n}\left(z\right)=\frac{\mathcal{K}_{3}\left(\frac{1}{z}\right)}{z\mathcal{K}_{2}\left(\frac{1}{z}\right)}-\left(4z+\frac{\mathcal{K}_{1}\left(\frac{1}{z}\right)}{\mathcal{K}_{2}\left(\frac{1}{z}\right)}\right)^{-1}\left(\frac{1}{z}+5\frac{\mathcal{K}_{3}\left(\frac{1}{z}\right)}{\mathcal{K}_{2}\left(\frac{1}{z}\right)}\right)\label{eq:ip}\end{equation}
As has been shown elsewhere \cite{ere}, in the non-relativistic limit,
$f_{T}\left(z\right)\rightarrow5/2$ and $f_{n}\left(z\right)\rightarrow0$.
Thus, the classical value $L_{T}=\frac{5}{2}\frac{nk^{2}T^{2}}{m}\tau$
is recovered for the thermal conductivity while the second term in
Eq. (\ref{eq:10}) vanishes rendering it a completely relativistic
effect \cite{L&L}.

The derivation of the relativistic hydrodynamic equations using Eq.
(\ref{eq:10}) as the constitutive equation for the heat flux is a
standard one and has been given in detail in various papers using
Fourier's equation \cite{JNT07,HL,Nos1}. The additional term in $\nabla n$
poses absolutely no problem at all, so we simply state the final form
for their linearized version namely,\begin{equation}
\delta\dot{n}+n_{0}\delta\theta=0\label{eq:18}\end{equation}
 \begin{eqnarray}
\frac{1}{c^{2}}\left(n_{0}\varepsilon_{0}+p_{0}\right)\delta\dot{u}_{\nu}+\frac{1}{\kappa_{T}}\delta n_{,\nu}+\frac{1}{\beta\kappa_{T}}\delta T_{,\nu}\nonumber \\
-\zeta\delta\theta_{,\nu}-2\eta\left(\delta\tau_{\nu}^{\mu}\right)_{;\mu}-\frac{L_{T}}{c^{2}}\delta\dot{T}_{,\nu}-\frac{L_{n}}{c^{2}}\delta\dot{n}_{,\nu} & =0\label{eq:19}\end{eqnarray}
 \begin{equation}
nC_{n}\delta\dot{T}+\left(\frac{T_{0}\beta}{\kappa_{T}}\right)\delta\theta-\left(L_{T}\delta T^{,k}+L_{n}\delta n^{,k}\right)_{;k}=0\label{eq:20}\end{equation}
where $\theta=u_{;\alpha}^{\alpha}$, $\zeta$ and $\eta$ are the
bulk and shear viscosity coefficients respectively, $\tau_{\nu}^{\mu}$
is the traceless symmetric part of the velocity gradient tensor, $\kappa_{T}$
the isothermal compressibility, $\beta$ the thermal expansion coefficient
and $C_{n}$ the specific heat at constant particle density. Nought
subscripts denote equilibrium quantities and $\delta\theta$, $\delta T$
and $\delta u_{\nu}$ denote the spontaneous fluctuations of the state
variables ($\delta\theta$ is related to $\delta n$ via Eq. (\ref{eq:18}))
around the equilibrium state of the fluid.

We remind the reader that this set of equations is the relativistic
analog of the one obtained from classical irreversible thermodynamics
to test the validity of the linear regression of fluctuations hypothesis
introduced by Onsager over seventyfive years ago \cite{OnsagerPR,Casimir}
as a basic ingredient required to derive his famous reciprocity relations.
The fact that this assumption has been experimentally tested for a
variety of systems \cite{deGrootJMP} calls immediately for its relevance
in the relativistic regime. To illustrate this point let us consider
the fluctuations associated to the transverse mode of the velocity
field $u^{\nu}$. This mode is easily uncoupled from the rest of the
fluctuations by taking the curl of Eq. (\ref{eq:19}). Since all the
gradient terms disappear we are led to the result that\begin{equation}
\frac{1}{c^{2}}\left(n_{0}\varepsilon_{0}+p_{0}\right)\delta\dot{U}_{\nu}-2\eta\nabla^{2}\delta U_{\nu}=0\label{eq:21}\end{equation}
where $\delta U_{j}=\epsilon_{jk}^{i}u_{;i}^{k}$ is the transverse
mode. Taking the Fourier transform of Eq. (\ref{eq:21}) we get that\begin{equation}
\delta\hat{U_{\nu}}\left(\vec{k},\, t\right)=\delta\hat{U_{\nu}}\left(\vec{k},\,0\right)\text{Exp}\left(-\frac{2\eta k^{2}c^{2}}{n_{0}\varepsilon_{0}+p_{0}}\right)\label{eq:22}\end{equation}
where $\delta\hat{U_{\nu}}\left(\vec{k},\, t\right)$ denotes the
Fourier transform of $\delta U_{\nu}$. Since the exponential's argument
is always negative, the fluctuations die out with a characteristic
time $\frac{\left(n_{0}\varepsilon_{0}+p_{0}\right)}{2\eta c^{2}k^{2}}$.
In the classical limit, since $\frac{n_{0}\varepsilon_{0}}{c^{2}}\rightarrow\rho_{0}$
and $\frac{p_{0}}{c^{2}}$ is negligible, the decay time reduces to
$\frac{\rho_{0}}{2\eta k^{2}}$ in agreement with classical hydrodynamics.
This result implies that the spontaneous fluctuations of such mode
obey the linear regression hypothesis. Notice that, since the last
two terms in Eq. (\ref{eq:19}) are precisely $\dot{J}_{[Q]}^{\ell}$,
this exponential decay is due to the fact that the curl of this vector
vanishes. Thus this result is independent of the equation of state
(see Eq. (\ref{eq:sand})).

From the remaining equations, Eq. (\ref{eq:18}), the divergence of
Eq. (\ref{eq:19}) and Eq. (\ref{eq:20}), one can get a set of coupled
equations for $\delta n$ and $\delta T$ whose solution is related
to the well known Rayleigh-Brillouin spectrum. A full discussion of
this feature is too long and technical to be presented here and will
be published elsewhere \cite{20}.

\section{Discussion and Concluding Remarks}

Examination of the results obtained in the previous section manifestly
indicate that Eq. (\ref{eq:10}) is the main result of this paper:
it satisfies all the requirements imposed by classical irreversible
thermodynamics and leads to a set of linearized hydrodynamic equations
which satisfy Onsager's linear regression of fluctuations hypothesis.
The crucial step was to eliminate the acceleration term appearing
in Eq. (\ref{eq:5}) in terms of the pressure gradient through Euler's
equations. This is a point that was overlooked by Eckart himself and
later by some of his followers and critics \cite{israel}-\cite{HL}.
Thus the term $\dot{u}_{\nu}$, which is not a force, has been replaced
by a combination of gradients of the state variables. This substitution
is necessary in order to arrive at a closed set of evolution equations
by further using the constitutive equations, which are written in
terms of such gradients. This is indeed the fundamental purpose of
the constitutive equations. If $\dot{u}_{\nu}$ is considered a state
variable the description of the system is overdetermined. This is
precisely the main source of the objections raised against Eckart's
theory which materialized in the appearance of the so called generic
instabilities \cite{HL,hl87} and presumably in giving rise to transport
equations of the parabolic type. This last aspect however deserves
a much closer attention and will not be addressed here. In fact, the
curious reader may easily verify that if $\dot{u}_{\alpha}$ in Eq.
(\ref{uno}) is expressed in terms of $\nabla p$ through Euler's
equation all the difficulties mentioned above will be eliminated.

In kinetic theory things are quite different. Although, as mentioned
in the text an acceleration term in Eq. (\ref{eq:5}) is present,
it is automatically disposed off when one appeals to the Chapman-Enskog
method for solving Boltzmann's equation. Indeed, when Eq. (\ref{eq:3})
is substituted in Eq. (\ref{eq:1}), the term to order zero in the
gradients yields Euler's equations. Successively, the equation for
the correction of order $n$ in the gradients will have a solution
if and only if the hydrodynamic equations of order $n-1$ in the gradients
hold true. This is the spirit of the original method of Hilbert to
solve Boltzmann's equation and this is why the drift term in the $n^{th}$
order equation which is already of order one in the gradients has
to be computed with the solution of order $n-1$. This explains why
the use of Euler's equations is demanded to calculate $\phi$ in Eq.
(\ref{eq:5}). We should also emphasize that these results have been
shown to hold for a relativistic Boltzmann equation with its full
collision term \cite{NOSPA} and further, from phenomenological arguments,
to be valid for any arbitrary relativistic fluid \cite{Nos1}.

Moreover, to fully grasp the relevance of this calculation we can
compare Eq. (\ref{eq:21}) with its counterpart obtained in Ref. \cite{HL},
in which the coupling of heat with acceleration is retained. In that
case, the transverse velocity mode fluctuations satisfy the equation:

\begin{equation}
\frac{\kappa T_{0}}{c^{4}}\delta\ddot{U}_{\nu}-\frac{1}{c^{2}}\left(n_{0}\varepsilon_{0}+p_{0}\right)
\delta\dot{U}_{\nu}+2\eta\nabla^{2}\delta U_{\nu}=0\label{eq:23}
\end{equation}

After taking the Fourier-Laplace transform in this equation, we now
obtain two roots governing the time behavior of the fluctuations.
The calculation in a specific example, water at 300 K at one atmosphere
leads to an extremely short characteristic time in the growing mode.
The conclusion included in that paper states that first order theories
should be discarded in favor of more complicated formalisms. Here
we have shown that the apparent trivial replacement of acceleration
with the pressure gradient while solving Boltzmann's equation, with
the Chapman-Enskog method, leads to a nice exponentially decaying
transverse velocity mode. Thus, the so-called generic instabilities
are indeed absent in first order relativistic hydrodynamics. In addition
to this application, we can assert that the calculation of the Rayleigh-Brillouin
spectrum using the acceleration term leads to the unphysical result
that such spectrum is inexistent. This has been fully discussed in
Ref. \cite{20}.

There is a last comment which is pertinent in view of the results
here presented. As we said in the introduction, Eckart's formalism
as well as all others which contain an entropy balance equation with
an entropy flux is given by $J_{[Q]}^{\ell}/T$, have been referred
to as {}``first order theories''. Due to the so-called generic instabilities
and the apparent violation of the principle of antecedence they where
substituted by second order theories \cite{israel-stewart,HL,NOSPA}
which postulate that both the entropy density as well as the entropy
flux should be functions of the heat flow and the viscous tensor.
The point here is to remind the reader that the first order theories,
if the above definition is kept, together with the ordinary linear
force-flux relations lead, for a simple viscous fluid in the non-relativistic
case, to the Navier-Stokes-Fourier equations of hydrodynamics. These
equations in their full expression consist of a set of five non-linear
coupled equations (first order in time derivatives and second order
in spatial derivatives). Their linearized version is in total agreement
with Onsager's regression assumption and only upon some drastic simplifications
lead to transport equations of the parabolic type, such as Fourier's
heat conduction equation. To assert that in the relativistic domain
such theories are incomplete and/or incorrect is equivalent to asserting
that there is no relativistic extension of the NSF equations. What
we have proved in this paper is that this requires further study.
What is the full and correct version of these equations may still
be today a problem which has not been completely solved but already
some attempts have been clearly discussed in the literature \cite{lcdspa}.

We believe that these considerations are important due to the enormous
effort that has been made in the past few years to describe the hydrodynamics
of the hot dense matter produced in heavy ion collisions. Some workers
in this field are using the model proposed by second order theories
or modified versions of it \cite{25koide}-\cite{31refprd} which
require the use of adjustable parameters \cite{32lgcs}. As an example
of another alternative to these theories there is Landau's hydrodynamics
for hot dense matter based on Euler's equations. In an elegant paper
by C. Y. Wong \cite{33wong}, this theory has been recently shown
to provide a very good explanation for the collective dynamics of
the fluid that is produced in the relativistic collisions of heavy
ions. If one would consider adding dissipative effects it is clear
that the NSF equations would be the best candidate. Thus we conclude
paraphrasing S. Weinberg's words {}``it is worth our while to develop
the outlines of the general theory of a relativistic imperfect fluid''
\cite{ww}.

\end{document}